\documentclass{ws-procs9x6}
\usepackage{graphicx}

\begin{document}

\title{Neutron Captures and the r-Process}

\author{T. RAUSCHER}

\address{Department of Physics and Astronomy,
University of Basel\\
Klingelbergstr.\ 82,
4056 Basel, Switzerland\\
E-mail: Thomas.Rauscher@unibas.ch}

\maketitle

\abstracts{The r--process involves neutron-rich nuclei far off stability
for which no experimental cross sections are known. Therefore, one has
to rely on theory. The difficulties in the predictions are briefly
addressed. To investigate the impact of altered rates, a comparison of
r--process production in hot bubble models with largely varied rates is
shown. Due to the (n,$\gamma$)-($\gamma$,n) equilibrium established at
the onset of the r-process, only late-time neutron captures are
important which mainly modify the abundances around the third r--process
peak.}

\section{Introduction}
Nucleosynthesis of elements beyond the iron peak requires reactions with
neutrons due to the high Coulomb barriers which prevent charged particle
reactions. Except for the relatively underabundant proton-rich p--nuclei,
two processes have been identified for the production of intermediate
and heavy nuclei: the slow neutron-capture process (s--process) and the
rapid neutron-capture process (r--process). With neutron number
densities around 10$^8$ cm$^{-3}$ and low effective neutron energies of
around 30 keV, the s--process synthesizes nuclei along the line of
stability as the neutron captures are generally slower than all
beta-decays encountered along its path (with the exception of several
branching points where the two timescales become similar). Approximately
half of the intermediate and heavy elements are created in the much
faster r--process with neutron number densities exceeding 10$^{22}$
cm$^{-3}$, effective neutron energies around 100 keV, and much shorter
process times of up to a few seconds. These conditions point to an explosive
site but the actual site has yet to be identified. The long favored idea
of a high-entropy bubble in the neutrino wind ejected from a type II
supernova shows persistent problems in explaining production
across the full mass range of r--nuclei. Furthermore, there are indications 
that there must
be two distinct sites ejecting r--process material at different
frequencies (see other contributions in this volume). In consequence, 
most r--process investigations focus on
simplified, parameterized models which allow to study the required
conditions and their sensitivities to nuclear inputs.

Due to the high neutron densities the r--process synthesizes very
neutron-rich nuclei far off stability which subsequently decay to
stability when the process ceases due to lack of neutrons or low
temperatures. This raises
the question whether we can predict reactions far off stability
sufficiently well to make statements about r--process conditions. In the
following two main topics are briefly addressed: The difficulties in
predicting neutron captures far off stability, and the impact of neutron
captures on the resulting r--process abundances.

\section{Predicting Neutron Capture}
\begin{figure}[htb]
%\epsfxsize=10cm   %width of figure - will enlarge/reduce the figures
%\epsfbox{fig3.eps}
%\figurebox{2cm}{3cm}{} %to have a box alone 
\centerline{\epsfxsize=4.1in\epsfbox{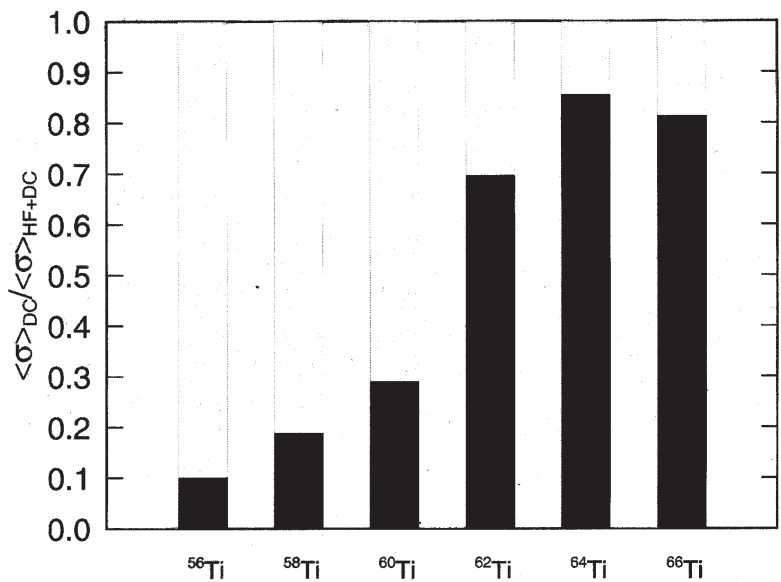}}   
\caption{Portion of direct capture in the (n,$\gamma$)
cross section for a series of
Ti isotopes from a comparison of Hauser-Feshbach and DC calculations.
Clearly, the DC contribution increases with decreasing neutron
separation energy. \label{fig:DCAnteil}}
\end{figure}

As the astrophysical reaction rate is obtained by folding the
energy-dependent cross section with the Maxwell-Boltzmann velocity
distribution of the projectiles, the relevant energy window for neutrons
is given by the location $E_0\approx 0.172 T_9 (\ell + 1/2)$ [MeV] and 
width $\Delta \approx 0.194 T_9 (\ell + 1/2)^{1/2}$ [MeV] of the maximum of the
Maxwell-Boltzmann distribution at the given stellar temperature. Since
the cross section is integrated over this energy window, the available
number of levels within determines the dominating reaction mechanisms.
With a sufficient number of overlapping resonances (about 10) the
statistical model (Hauser-Feshbach) can be used which employs averaged
transmission coefficients and describes the reaction proceeding via a
compound nucleus [1]. Single, strong resonances destroy the notion of the
simple energy window as the integrand is split in several terms.
Finally, in between resonances or without resonances, direct capture
will become important. The temperatures above which the statistical
model is applicable for the calculation of neutron- and charged-particle
induced reaction rates have been estimated in [2]. Explicit
limits are given in the global calculation of statistical model rates of
[1]. These limits should be taken as a guideline when applying
the rates given therein. Fig.\ \ref{fig:DCAnteil} shows how direct
capture becomes more and more important for nuclei with lower and lower
neutron separation energy.

Basically, there are three groups of problems connected to the
prediction of rates far from stability. The first two (partially
overlapping) groups concern the
difficulty in predicting nuclear properties relevant for Hauser-Feshbach
and direct capture. 
For more details on these, see, e.g., [3]. Here, only the most important
topics are outlined.

Direct capture calculations are extremely sensitive to the nuclear
input, such as neutron separation energies, spins, parities and
excitation energy of low-lying states, and the potential used in the
neutron channel [4]. One of the largest problems is the determination of
the spectroscopic factor which is difficult to calculate. At stability
it is usually derived from (d,p) data. However, even there a
considerable uncertainty is involved as it is taken from a comparison of
prediction and data and thus is not independent of theory.

Due to the nature of the statistical model and its use of average
quantities its sensitivity to most nuclear inputs is not as extreme as in
the direct capture case. Nevertheless, it is yet uncertain how well the
relevant nuclear properties, such as the particle separation energies,
neutron optical potential,
level density, and the low-energy tail of the GDR, can be described far
off stability. Global models, in which the properties are not optimized
to a few nuclei or a single mass region but rather are attempted to be
consistently predicted for all nuclei, fare very well along stability.
However, since the used descriptions are derived from data at stability
(by either adjusting phenomenological or microscopic parameters) it
remains an interesting question whether they are still valid far off
stability. Nevertheless, as pointed out above, the statistical model
is not applicable at low neutron separation energies and therefore the
impact of the uncertainties far off stability are limited.

The third problem is the identification of the dominant
reaction mechanism and the interplay of different reaction mechanisms
when their contributions are of similar size. Clearly, more work has to
be done on this in the future. Lacking other data, basically all
astrophysical investigations use Hauser-Feshbach rates even for isotopes
where it is not applicable. With a low level density it is usually
expected that the statistical model overestimates the actual cross
section, unless strong, wide resonances are found in the relevant energy
window.

\section{Implementation of Neutron Capture in the r-Process}

\subsection{General}
\begin{figure}[htb]
%\epsfxsize=10cm   %width of figure - will enlarge/reduce the figures
%\epsfbox{fig3.eps}
%\figurebox{2cm}{3cm}{} %to have a box alone 
\includegraphics[width=8cm,angle=-90]{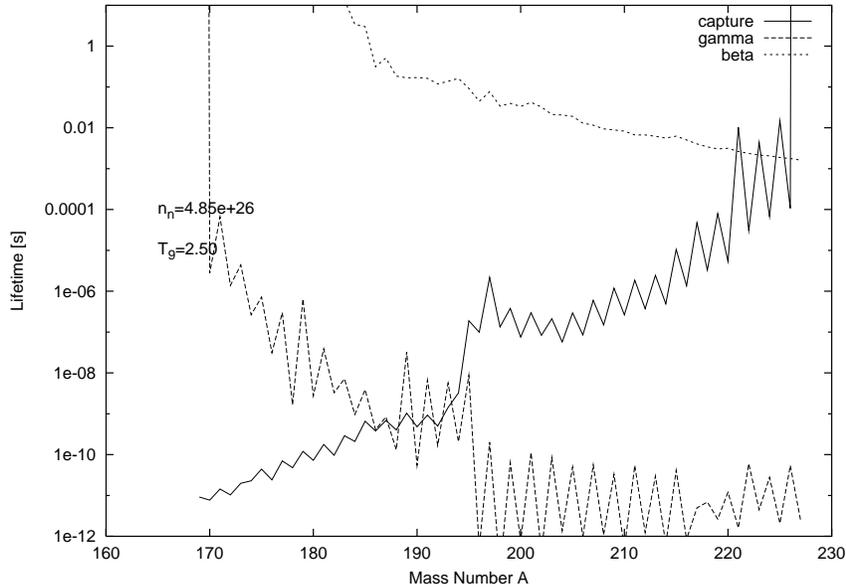}
%\centerline{\epsfxsize=4.1in\epsfbox{times150.ps}}   
\caption{Lifetimes against (n,$\gamma$) (full line), ($\gamma$,n)
(dashed), and $\beta$-decay (dotted) of neutron-rich Tm isotopes. Captures
and photodisintegrations are much faster than $\beta$-decays and
abundances are determined by an (n,$\gamma$)-($\gamma$,n) equilibrium.
(Lifetimes at the edges of the considered chain have been set to high
values to prevent mass loss from the network.)
\label{fig:lifetimes200}}
\end{figure}

Given the difficulties in predicting rates far off stability, one might
wonder whether it is possible at all to study the r--process, even if
one resorts to simply parameterized networks. However, the situation is
not that bad since it is not necessary to know the rates directly in the
r--process path. Contrary to a sometimes still persisting misconception,
the formation of r--isotopes {\it cannot} be viewed as occurring by a sequence
of neutron captures until reaching an isotope with a $\beta$--lifetime
shorter than the neutron-capture lifetime, somewhat like an s--process
but moving further out from stability. As shown in Fig.\
\ref{fig:lifetimes200}, all neutron captures and
photodisintegrations occur faster by several orders of magnitude than
any $\beta$-decay in a given isotopic chain. In fact, the reactions are
so fast that almost instantaneously ($<10^{-8}$ s)
an equilibrium state is reached in
which the abundance $Y$ for each isotope is determined by the balance of
the reactions creating and destroying it: 
$r_\mathrm{(n,\gamma)}Y_A=r_\mathrm{(\gamma,n)}Y_{A+1}$.
Since the two rates are related by detailed balance, the cross sections cancel
out and the ratio is mainly depending on $S_\mathrm{n}$, $T$, and
$\rho$. Neutron captures will only start to matter during freeze-out
when the lifetimes become longer due to lower temperatures and lower
neutron number densities. It has been shown that the
freeze-out proceeds very quickly for realistic conditions [5].
On one hand
this limits the importance of neutron captures, on the other hand it
validates the investigations which were performed using approximations
such as instantaneous freeze-out [6].
\begin{figure}[htb]
%\epsfxsize=10cm   %width of figure - will enlarge/reduce the figures
%\epsfbox{fig3.eps}
%\figurebox{2cm}{3cm}{} %to have a box alone 
\includegraphics[width=8cm,angle=-90]{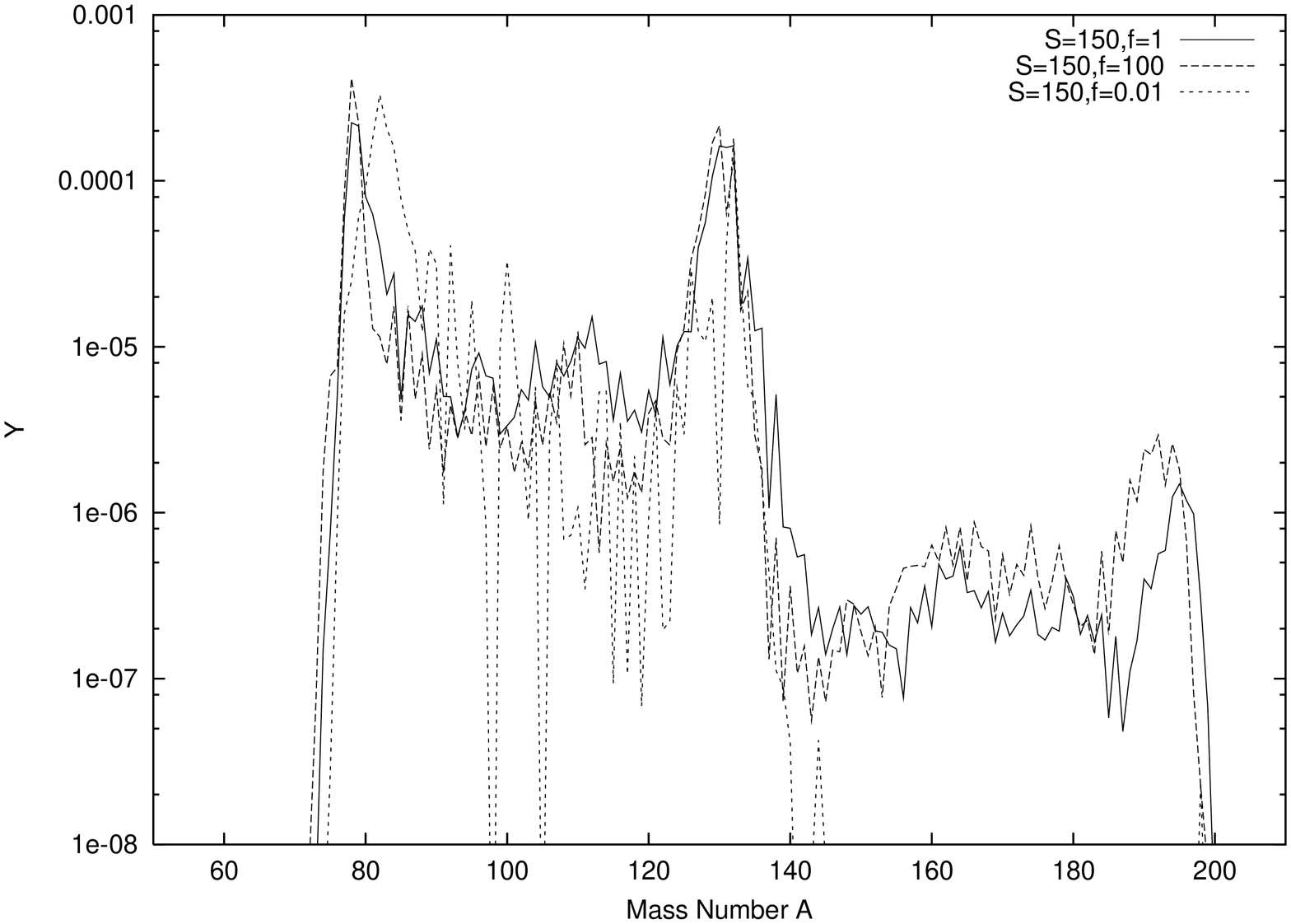}
\caption{Decayed final abundances of the $S=150$ models. The neutron
rates were multiplied by factors 1.0 (full line), 100. (dashed), and
0.01 (dotted), respectively.
\label{fig:s200abun}}
\end{figure}
\begin{figure}[htb]
%\epsfxsize=10cm   %width of figure - will enlarge/reduce the figures
%\epsfbox{fig3.eps}
%\figurebox{2cm}{3cm}{} %to have a box alone 
\includegraphics[width=8cm,angle=-90]{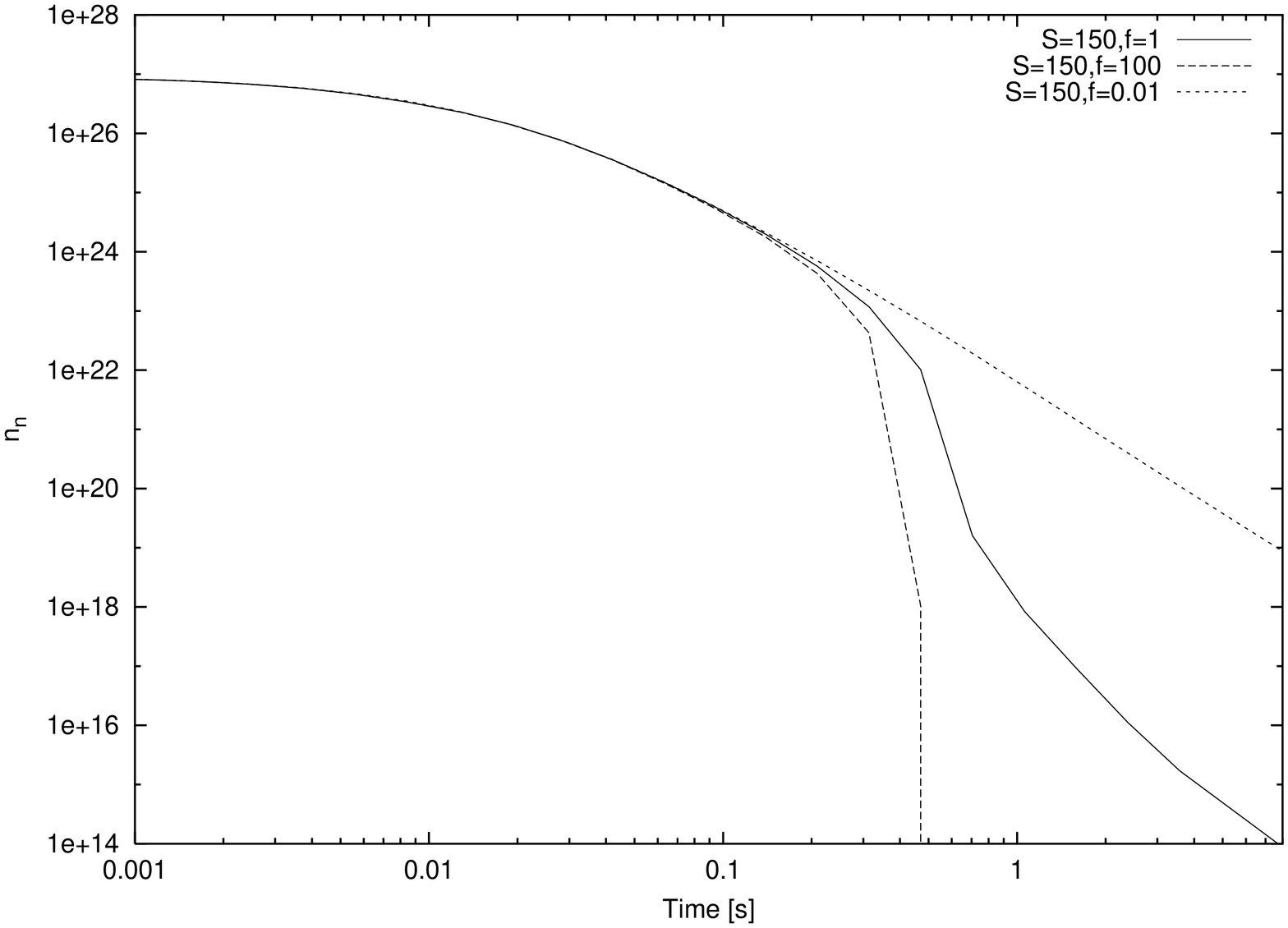}
%\centerline{\epsfxsize=4.1in\epsfbox{DCAnteil.eps}}   
\caption{Time evolution of the neutron number density in the $S=150$
models. The neutron rates were multiplied by factors 1.0 (full line),
100. (dashed), and 0.01 (dotted), respectively.
\label{fig:s200nn}}
\end{figure}
\begin{figure}[htb]
%\epsfxsize=10cm   %width of figure - will enlarge/reduce the figures
%\epsfbox{fig3.eps}
%\figurebox{2cm}{3cm}{} %to have a box alone 
\includegraphics[width=8cm,angle=-90]{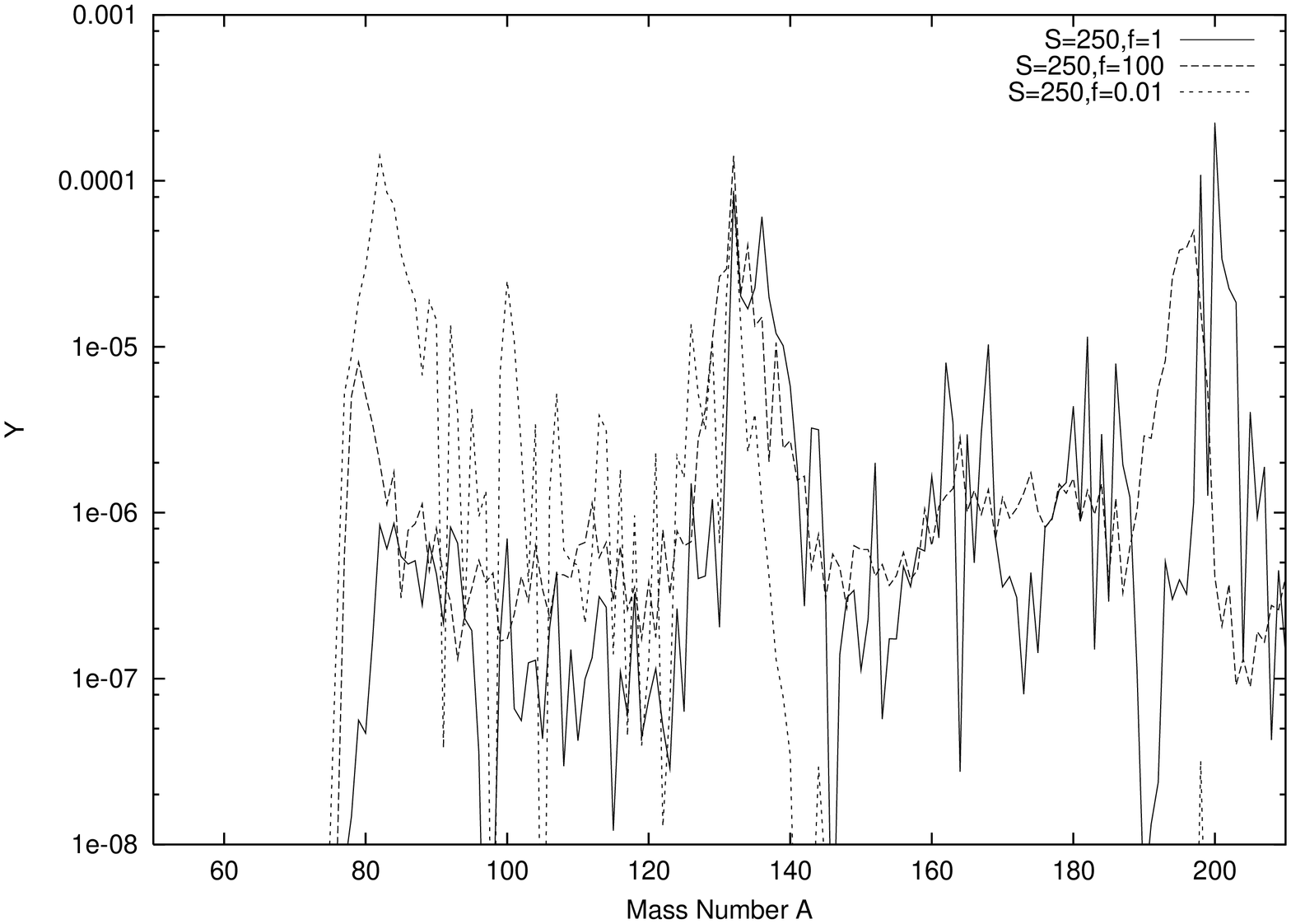}
%\centerline{\epsfxsize=4.1in\epsfbox{DCAnteil.eps}}   
\caption{Decayed final abundances of the $S=250$ models. The neutron
rates were multiplied by factors 1.0 (full line), 100. (dashed), and
0.01 (dotted), respectively.
\label{fig:s300abun}}
\end{figure}
\begin{figure}[htb]
%\epsfxsize=10cm   %width of figure - will enlarge/reduce the figures
%\epsfbox{fig3.eps}
%\figurebox{2cm}{3cm}{} %to have a box alone 
\includegraphics[width=8cm,angle=-90]{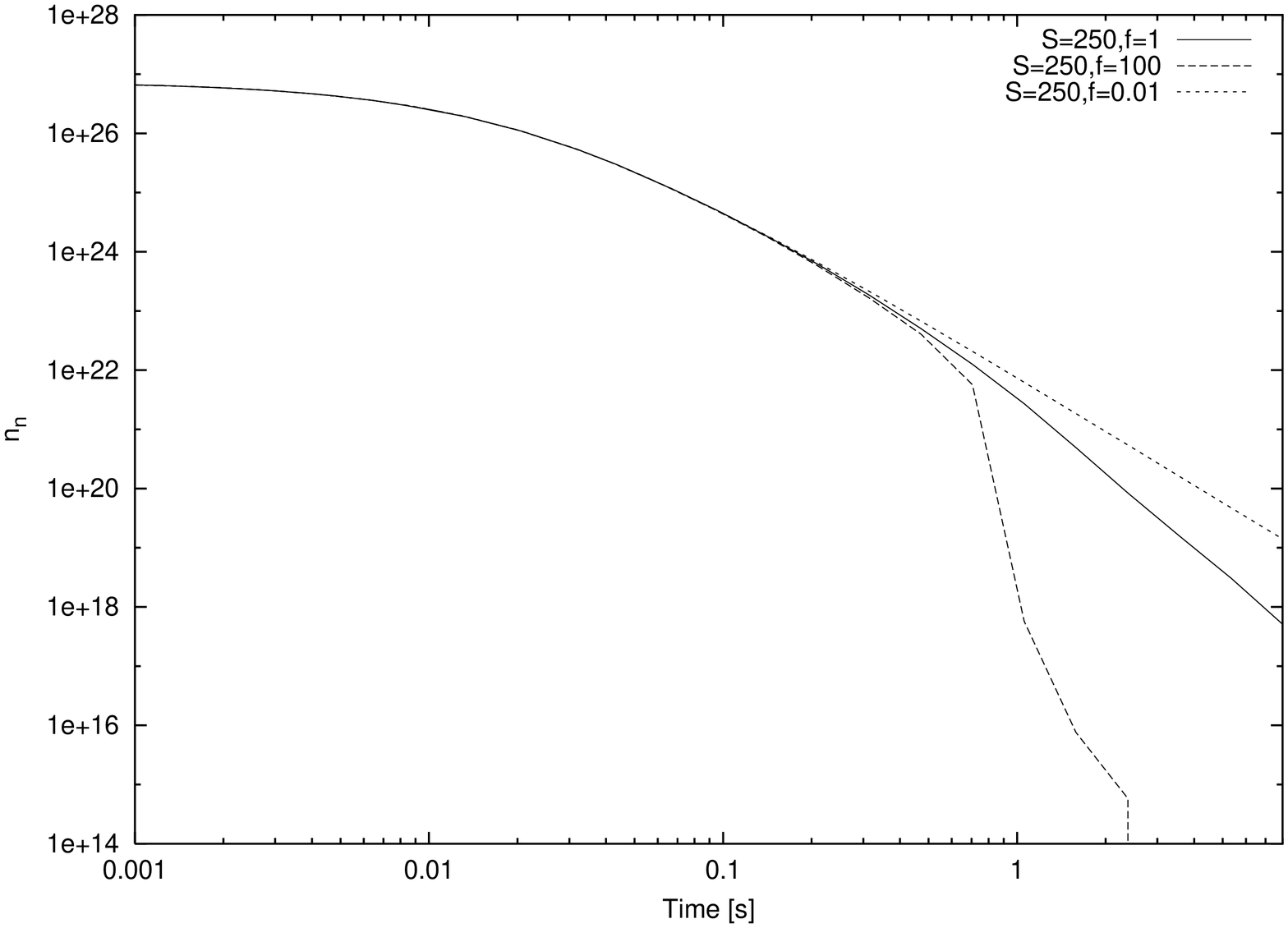}
%\centerline{\epsfxsize=4.1in\epsfbox{DCAnteil.eps}}   
\caption{Time evolution of the neutron number density in the $S=250$
models. The neutron rates were multiplied by factors 1.0 (full line),
100. (dashed), and 0.01 (dotted), respectively.
\label{fig:s300nn}}
\end{figure}

\subsection{Dynamic r-process simulations}
In order to study neutron captures in the freeze-out it is necessary to
perform dynamic r--process simulations. As an example,
calculations in the model of an adiabatically expanding hot bubble were
performed,
similar to [5] but with updated, temperature-dependent rates, including
the theoretical rates of [1].
In this model of a primary r--process, a blob of matter at high
temperature ($T_9\approx 9$) expands and cools. For the calculations here
the same expansion was chosen as used by [5] in their case of 50
ms expansion timescale. Due to the initial high temperature, all
reactions, including charged-particle reactions, are in equilibrium and
the resulting abundances can be calculated for each temperature from the
equations describing a full NSE. The charged-particle reactions, in
particular the $\alpha$ captures, cease at around $T_9 \approx 2.5$.
Below that temperature it is not necessary to use a full network but one
can utilize a simpler network, only including (n,$\gamma$),
($\gamma$,n), and $\beta$--decays. The seed abundances for this
r--process network are given by the freeze-out abundances of the charged
particle network. More specifically, depending on the freeze-out
conditions the slow triple--$\alpha$ rate will either be able to convert
all $\alpha$'s to heavy mass nuclei or it will be too slow, leaving a
certain $\alpha$ mass fraction. The latter is called $\alpha$-rich
freeze-out. The process conditions are specified by the entropy $S$, the
electron abundance $Y_\mathrm{e}$, and the expansion timescale. Depending on the
conditions, more or less free neutrons per heavy seed nucleus
are available after the
charge-particle freeze-out. Due to the still high temperature an
(n,$\gamma$)-($\gamma$,n) equilibrium is established. The
$\beta$-halflife of the most abundant nuclei in each isotopic chain
(these are only one or two due to the shape of the equilibrium equation)
determine how fast material can be converted to the next element. Each
chain remains in equilibrium until finally the r--process freezes out.

For this  comparative study, representative seed abundances were chosen
without following the full charged particle network. The calculations
always started by only populating the Fe chain but with realistic
$Y_\mathrm{n}/Y_\mathrm{seed}$ and $Y_\alpha$ (depending on entropy and
$Y_\mathrm{e}$) taken from parameterized results of full calculations.
Since the uncertainties in the neutron capture rates might be large, for
two entropies 3 exemplary cases are shown here: with standard rates
and with neutron captures multiplied by a factor of 100 and a factor of 0.01,
respectively (this implies that the photodisintegrations are changed by
the same factor). Figs.\ \ref{fig:s200abun} and \ref{fig:s300abun} show the
final abundances, the neutron number densities as a function of time are
shown in Figs.\ \ref{fig:s200nn}, \ref{fig:s300nn}. At low entropy there
are not enough free neutrons to considerably change the seed abundance,
the neutron freeze-out is also fast. It was already shown in
[5] that the freeze-out at higher entropy is slower and that
final neutron captures can alter the resulting abundances of heavy
nuclei but not of light ones. The trough before the high-mass peak
was filled by late neutron captures.

The freeze-out behavior obtained here depends
on the chosen neutron rates. The time at which the $n_\mathrm{n}$ for
the three cases diverge indicates the fall-out from the rate
equilibrium. After this point it depends on the entropy how far up in
mass nuclei have been produced and on the neutron captures how their
abundances are altered. As can be seen in Figs.\ \ref{fig:s200nn} and
\ref{fig:s300nn}, the final freeze-out phase is faster for larger rates.
This reflects the increased capture when the forward and reverse rates
fall out of equilibrium which uses up neutrons faster. The masses above
about 140 are mainly produced in this late freeze-out phase and are
therefore more sensitive to the value of the neutron captures.
Especially in the high entropy case shown in Fig.\ \ref{fig:s300abun} it is
evident that faster neutron captures smooth the abundance distribution
and fill the trough before the $A\approx 200$ peak. For both entropies,
the artificially suppressed rates do not allow to build up considerable
abundances beyond $A\approx 140$.

\section{Conclusion}
The simple comparison shown above for the hot bubble model has to be
interpreted cautiously. Despite the fact that there might be
considerable uncertainties in the theoretical rates far off stability
changing {\it all} rates in a range of 4 orders of magnitude seems
unrealistic. Even if new effects (like pygmy resonances [7] or
overestimated cross sections [3])
might change the rates by factors of 10 for extremely neutron-rich
nuclei, late-time captures will not include such nuclei but will occur
closer to stability. Moreover, for reproducing the solar r--process
pattern it is necessary to superpose a number of components with
different entropies. Thus, effects of rates altered on a large scale, as
shown above, can be compensated by a scaling in entropy and a different
weight distribution. Thirdly, a more realistic seed abundance
distribution might also decrease the difference in heavy element
production between the different cases. Higher entropies
realistically start with seed abundances in the $A\approx 110$ region
and require less neutrons to form more heavy elements. However, this was
not taken into account here to purely show the influence of altered
neutron captures.

Despite the above caveats the main conclusions are consistent with other
studies [5,8].
Components with high entropy freeze out slower and late-time
neutron captures can modify the final abundance distribution mainly in
the region $A>140$. Therefore, emphasis has to be put on improving the
prediction of nuclear cross sections and astrophysical reaction rates in
that mass region.

\section*{Acknowledgments}
This work was supported by the Swiss NSF, grant 2000-061031.02. T. R.
acknowledges a PROFIL professorship of the Swiss NSF (grant
2024-067428.01).

\end{document}